# Efficient, self-phase matched, narrow-band, engineered shape, tunable high-order harmonic generation based on femtosecond dynamic grating

Leila Mashhadi and Gholamreza Shayeganrad

**Abstract:**

The high-order harmonic generation (HHG), if phase matched, opens a unique way to produce coherent ultrafast extreme-UV (XUV) or soft x-ray light sources. Here we describe a robust and tunable method for scaling and narrowing the bandwidth of high-order harmonics in different transversal modes based on femtosecond dynamic grating in a gas jet. This technique has potential of self-phase controlled XUV generation by adjusting the intersect angle and the intensity ratio of interfering fundamental beams. The limitations of using high intensity laser fundamental pulses in gas, such as non-negligible the magnetic field which prevents recombination with parent and self-focusing of laser and plasma creation in the generated harmonics beam path which prevents phase matching, can be solved by using dynamic grating generated by the fundamental pulse and/or its harmonic with small intersect angle. The grating pattern introduces a phase modulation in the dipole emission imprint such pattern in the gas medium. As the spatial and temporal behaviors of the grating are closely connected in harmonic generation, by shaping of the spatial profile of the interfering beams it is possible to create a more complex tunable beating pattern as harmonic sources to transfer desired optical information to XUV pulses. In this case, the phase mismatch can be completely compensate by tuning the gas pressure, adjusting the intersect angle as well as intensity ratio of the interfering beams to produce a bright XUV with desired transversal mode. The residual phase mismatch in the harmonics can be compensate with the extra phase mismatch which is induced by noncollinear interference depends on the interfering beam spatial characterization and the intersect angle. The most impressive advantage of this technique is using mid-IR driving laser in perfect phase matched hard X-ray generation where the long wavelength driving laser dramatically decreases the HHG intensity.

1. Introduction

So far, most of the investigations of high-order harmonic generation (HHG) from gaseous media has focused on production of harmonic radiation using one driving laser beam. In such process an intense femtosecond laser pulse is focused in a gas jet to generate XUV or soft X-ray radiation collinearly with respect to the driving laser beam [1]. The conversion efficiency of this process can be enhanced by using an intense laser beam or increasing gas pressure. However, in both of them generation plasma and self-focusing disrupts phase matching condition between fundamental and harmonics. With the purpose of improving the phase matching counter propagating pulse has been used with driving laser beam. The electric field of backward travelling pulse when added to the driving pulse induces a direct and indirect phase modulations to the harmonics [2]. In other cases, two-color HHG driven by an IR laser and its second harmonic provides sub-cycle control of the generation electric field which has application to generate even and odd harmonics with enhancing conversion efficiency [3], and production of attosecond pulse trains with single cycle IR pulse [4]. Recently, generation of XUV optical vortex is implemented by interfering a strong Gaussian driving pulse and a week Laguerre Gaussian perturbing one [5]. In other work, noncollinear geometry used to generate angularly separated circularly polarized high harmonics [6]. By controlling the electron wavepacket recollision by interfering fundamental driving field with its detuned second harmonic, attosecond isolated pulses have been observed [7]. In other work, using two-color orthogonally polarized driving IR pulses the high harmonic signal is shown to be increased [8]. In these works and also in similar papers, the effect of transient gratings produced by interfering of two femtosecond pulses of same or different wavelengths considering the transversal variation of intensity on the generation of phase matched high-order harmonics narrow band with spatial transversal modes is not investigated.

Here we show that how a dynamic grating which is generated by the interference of two coherent femtosecond IR/visible or UV pulses can be used to control the intensity distribution and phase of the produced XUV pulses which has application in microscopy down to nanometric scale and manipulating the detecting object with complex rotato-vibrational moving. The effect of moving grating generated by the fundamental pulse and its harmonics in phase matching condition is investigated. It is shown that the residual phase mismatch in the harmonics can be compensate with the extra phase mismatch which is induced by noncollinear interference depends on the interfering beam spatial characterization and the intersect angle. The effect of dynamic grating created by Gaussian or LG beam, same frequency or harmonic, same intensity or perturbation one can be seen in the temporal and spatial properties of the harmonics.

HHG is achieved by focusing an intense femtosecond IR field into a gas target. The target, either atomic or molecular, is commonly found in experiments as a gas jet, a gas cell, or a gas filled capillary. The high nonlinear interaction between the IR field and each atom results in the emission of harmonics of the fundamental IR field whose frequency extends into the XUV or even the soft X-ray regime. Notably, radiation emitted by the atoms is coherent. Hence, the harmonic signal reaching the detector is strongly affected by the phase matching of the high-order harmonics emitted by each atom of the target. Thus, HHG radiation results from the interplay between the microscopic single-atom emission and the macroscopic superposition of the contributions of all the atoms of the target. From the microscopic point of view, the single-atom HHG process is composed of three step: ionization which happens likely near the peak of laser subcycle, propagation in continuum and photo-recombination. In detail, First, at some initial time $t_i$, close to the peak of the laser electric field, an electron wavepacket tunnels through the potential barrier formed by the combined Coulomb and laser fields; next, it oscillates almost freely in the laser field, gaining kinetic energy; finally, this energy is converted into a high-energy photon through recombination with the parent ion at time $t_r$. Each step contribute an amplitude and phase to the emitted field [9]. An infrared pulse lasting a few cycles allows the production of a single x-ray burst with a continuous spectrum. Using a multicycle laser pulse, this sequence is repeated every half cycle of the laser optical field, leading to periodic emission of light bursts with a discrete spectrum containing only odd multiples of the laser frequency: $\omega_q = q\omega_0$, where $\omega_0$ is the fundamental laser angular frequency, and q is an integer number so called the harmonic order. The temporal profile of the electric field or intensity of the emitted pulses can be determined from the knowledge of the spectral amplitudes and phases by Fourier transform integral. In macroscopic point of view, atoms located at different positions in the target emit harmonic radiation whose phase depends on the amplitude and phase of the local driving field. Then femtosecond transient grating as a driving source directly changes harmonic phase matching condition and creates the spatial regions in which the desired harmonics contribute efficiently and consequently can be developed as unique sources such as soft Xray harmonics driven by mid-IR interfering lasers beams [10]. In Section 2, we discuss in detail femtosecond transient grating. Section 3 specifically turns to the engineered shape HHG generation using transient grating and phase-matching condition.

## 2. Interference of two femtosecond pulses

To make a transient grating the first point is that the two light beams should be mutually coherent. We consider the general case that the two coherent light pulses with wavelength $\lambda_1$ and $\lambda_2$ and polarization $\mathbf{e}_1$ and $\mathbf{e}_2$ are propagating in x-z plane and intersect at an angle θ. The electric field of two pulses in paraxial approximation can be written as:

$$\mathbf{E}_{1,2}(\mathbf{r}_{1,2},t) = \mathbf{e}_{1,2} E_{01,2}(r_{\perp 1,2}) f_{1,2}(t) \exp\left\{-i\left[\omega_{1,2}t + k_{1,2}z_{1,2} + \Phi_{1,2}(t)\right]\right\} \quad (1)$$

where $f_{1,2}(t) = \exp\left[-2\ln 2(t/\tau_{1,2})^2\right]$ are envelope functions with Gaussian profile and $\omega_{1,2}$, $k_{1,2}$ and $\Phi_{1,2}$ are the carrier frequency, wave number and envelope phase which stands for the chirp of the carrier frequency of the interfering pulses, respectively. Here two beams are presented in local coordinates $(r_{\perp 1,2}, z_{1,2})$ where $r_{\perp 1,2} = r_{1,2} \cos\varphi_{1,2}$ with $\varphi_{1,2}$ as azimuthal angle in respective local coordinate. By transferring to (x,y,z) coordinate (see Fig. 1), we have:

$$\begin{pmatrix} x_{1,2} \\ y_{1,2} \\ z_{1,2} \end{pmatrix} = \begin{pmatrix} \cos\theta_{1,2} & 0 & \sin\theta_{1,2} \\ 0 & 1 & 0 \\ -\sin\theta_{1,2} & 0 & \cos\theta_{1,2} \end{pmatrix} \begin{pmatrix} x \\ y \\ z \end{pmatrix}. \qquad (2)$$

Assuming that there is no delay between two pulses and also $\Phi_1(t) = \Phi_2(t)$, the intensity distribution of the total field in space-time gives a dynamic pattern which can be written as

$$I(\mathbf{r},t) = \mathbf{E}_{tot}(\mathbf{r},t) \cdot \mathbf{E}^*_{tot}(\mathbf{r},t) = E^2_{01}(r) f_1^2(t) + E^2_{02}(r) f_2^2(t) + 2 E_{01}(r) E_{02}(r) f_1(t) f_2(t) \cos(\delta) \qquad (3)$$

where $\delta$ is the angle between electric field of the beams, $\delta = \Delta\mathbf{k}\cdot\mathbf{z} - \Omega t$ with $\Delta\mathbf{k} = \mathbf{k}_2 - \mathbf{k}_1$ and $\Omega = \omega_2 - \omega_1$. In this condition, the interfering beams make an intensity modulation which depends on both x and z coordinates.

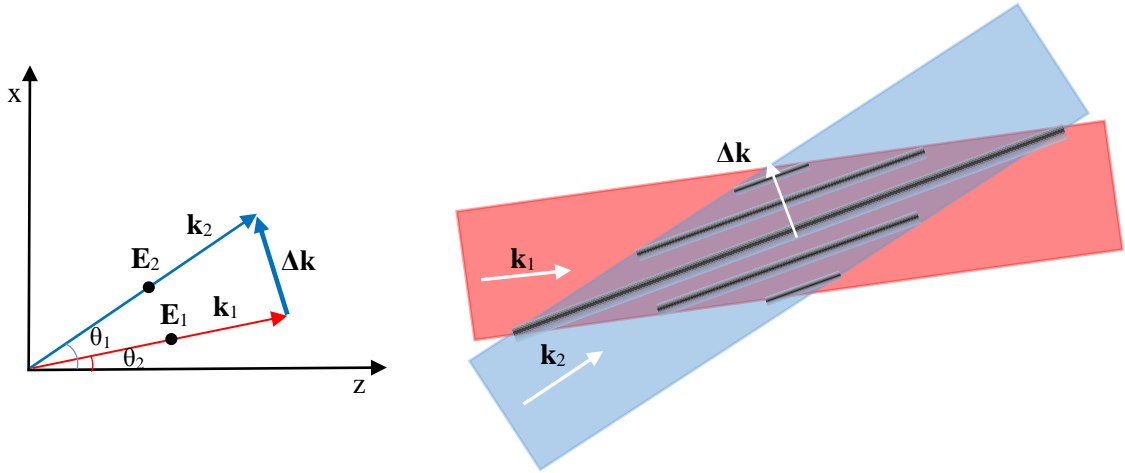

Fig. 1: Schematic of interference of two beams with different frequency and wave vectors propagating in local coordinates ($x_{1,2} y_{1,2} z_{1,2}$). The fringe pattern produced by two beams moves in the direction of $\Delta\mathbf{k}$.

The polarization of the interfering pulses is important since it can alter the modulation depth of the interference pattern. Here, we supposed that the polarization of both beams are along y-direction. According to Eq. (3), if two beams has perpendicular polarization, no interference pattern will appear.

Fro Eq. (3), in the case of interfering two beams with different wavelength, the interference pattern will move with velocity defined by:

$$V = \frac{\Omega}{|\Delta\mathbf{k}|} \qquad (4)$$

Considering, for instance, 800 nm fundamental pulse and a fraction of its second harmonic pulse which intersect at angle 4°, the produced fringes moves with velocity which be almost equal to the velocity of light (c=300 nm/fs). Fig. 2 shows the periodicity of the grating pattern versus intersect angle and detuning wavelength of the interfering beams. It can be seen that the grating period and moving velocity of the pattern reduces with intersect angle. In addition, at a fixed intersect angle, the grating periodicity decreases with detuning wavelength, while the moving velocity of the pattern increases very fast with detuning wavelength and this trend gets slower and for $\Delta\lambda>0.4$ nm it reached velocity of light. It is obvious that the grating is stationary when the two beams have the same frequency. Note that the nonlinear response of atoms are fast enough to experience the traveling of the fringe pattern.

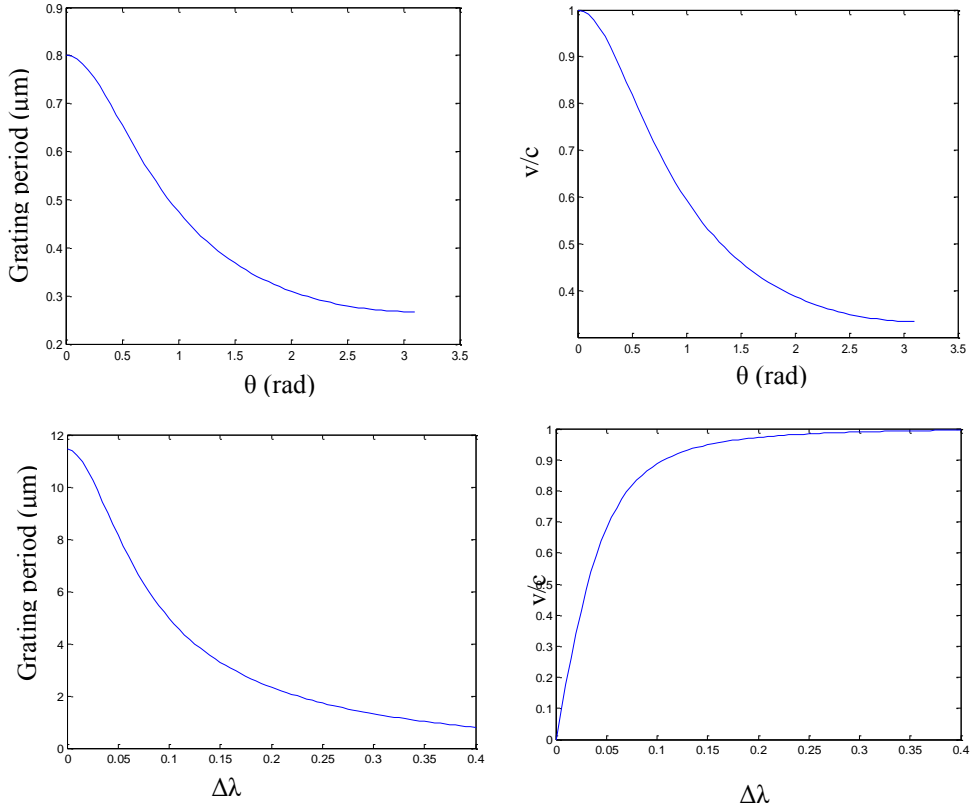

Fig. 2. Periodicity and moving velocity of the grating produced by two beams with wavelength 800 nm and 400 nm as a function of intersect angle (up) and wavelength detuning at a fixed intersect angle 4° (down).

The modulation depth visibility or the energy density of the grating per unit area is given by [11]:

$$\Xi(\mathbf{r}) = \int_{-\infty}^{\infty} I(\mathbf{r},t)dt \qquad (5)$$

If the pulses have the duration of $\tau_1$ and $\tau_2$ and frequency detuning $\Omega=\omega_1-\omega_2$, the modulation depth visibility is given by:

$$\Xi(\mathbf{r}) = \sqrt{\frac{\pi}{4\ln 2}} \left( E_{01}(\mathbf{r})^2 .\tau_1 + E_{02}(\mathbf{r})^2 .\tau_2 + \frac{2E_{01}(r_\perp)E_{02}(r_\perp)}{\sqrt{\frac{1}{\tau_1^2}+\frac{1}{\tau_2^2}}} e^{-\frac{\Omega^2 \tau_1^2 \tau_2^2}{8(\tau_1^2+\tau_2^2)\ln 2}} \cos(\Delta \mathbf{k}.z) \right) \qquad (6)$$

For two pulses with the same duration $\tau_1 = \tau_2 = \tau$ the average of modulation depth versus the pulse duration frequency detuning is shown in Fig. 3. According to Fig. 3, for short pulses there is a possibility to produce moving grating with high modulation depth. However, the deepest modulation is achieved when grating is stationary.

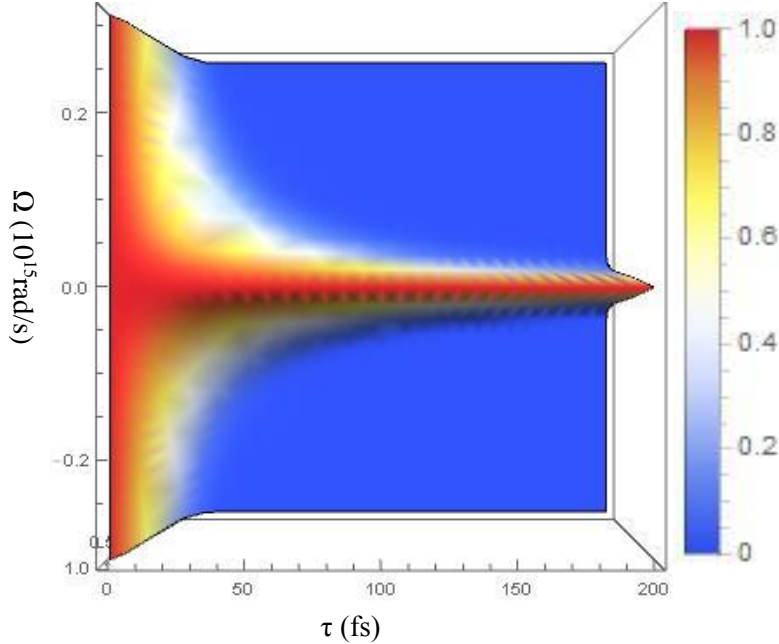

Fig. 3. Modulation depth as a function of frequency detuning $\Omega$ and pulse duration $\tau$. The deepest modulation is achieved when grating is stationary.

### 3. HHG based on dynamic grating

In HHG based on femtosecond transient grating a fundamental linearly polarized femtosecond Gaussian laser beam (e.g. 800 nm from Ti:sapphire system) with intensity of the order of $10^{14}$ W/cm$^2$ and an interfering pulse with controllable transversal shape and polarization are superimposed at tunable angle. pulse duration of few 10 fs. The duration of the pulses should fulfill the criteria of the modulation depth according to the Fig.3. The temporal overlap of two pulses are defined by a translation stage in the perturbing beam path (see Fig. 4). By introducing a desired beam shaper in the interfering beam path, the transversal mode will be engineered to create a desired transversal shape. For example by passing the perturbing beam trough a q phase plate it can be converted to Laguerre Gaussian (LG) mode with an orbital angular momentum [11]. The polarization is controlled by introducing birefringence retarding plates. Both beams are focused to the estimated intensity ratio $I_1/I_2$ at the intersect angle in a noble gas medium. The superposed coherent driving pulses and the interfering pattern yields a spatially modulated distribution of the energy density. As mentioned above, the grating pattern is static if both beams have same color, whereas the moving grating can be produced by the detuning wavelength of the interfering beams or by interfering the fundamental beam with its harmonic (for instance second its second which is generated in BBO as shown in Fig. 4). The interaction of the created dynamic grating with atoms imposes a suitable grating shape for desired harmonic mode onto the plane of emitting dipoles which leads to create a periodic-sources for HHG of the fundament which carrying the information of interfering beam. By adjusting the intersect angle periodicity of the grating is defined for constructive interference of the HHG sources generated by the dynamic grating. As explained in section 2, the period and the velocity of the grating can be tuned by adjusting the intersect angle and wavelength detuning.

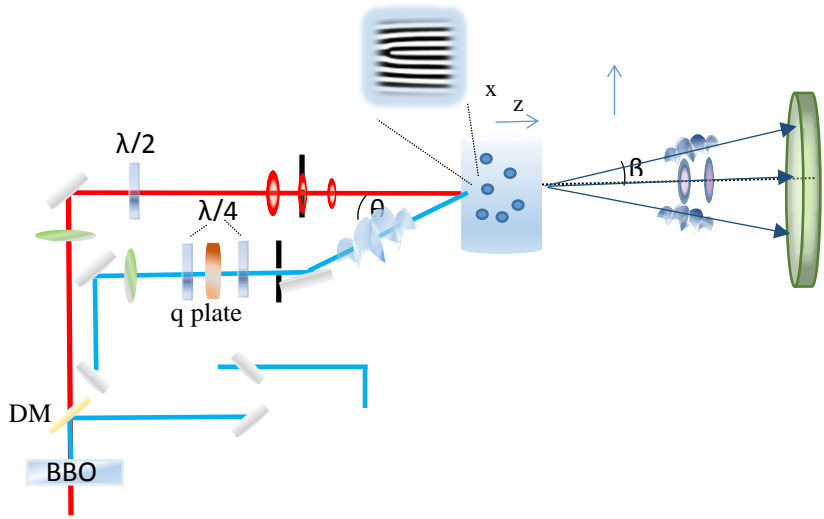

Fig. 4. Schematic of possible optical setup for engineered transversal mode HHG by moving grating. 800 nm collimated beam passing through BBO crystal to generate 400 nm. The optical path of the beam and its second harmonic are matched by using an optical delay line in the interfering beam path. Half wave plate ($\lambda/2$) and quarter wave plate ($\lambda/4$) are used to control the polarization of interfering beams. The q plate converts the Gaussian beam into LG mode. Irises are used for tuning the intensity. Lenses in the beam path can be used to optimize the focusing geometry. The pulses are superimposed at the angle θ (a few degree) in gas jet. The grating source in rare gas generates multiple beamlets of different order of transversal modes in the emission cone that can be recorded by a MCP and/or phosphor screen for position-sensitive detection. By removing BBO crystal from the setup, same setup can be used for HHG generation in statistic grating of two beams having similar wavelengths.

Therefor a spatiotemporal interference of HHG will be achieved which increases the HHG intensity by a factor that is depended on the number of bright points of driving laser pulse within interacting area. Adjusting interference of pulses should result in shorter XUV pulses. To synchronize perfectly the generated pulses in it possible to use diffraction pattern of two LG beams with opposite angular momentum. This pattern contains several maximum loops on a circle in the transversal plane respect to the propagation direction.

The produced moving grating can be created by Pulses with different transversal modes to imprint also the desired transversal shape on the generated harmonics with high conversion efficiency. The velocity of the grating and the period can be controlled by wavelength detuning as well as by intersect angle. This technique can be used to produce phase matched XUV in desired transversal modes by taking the advantage of XUV diffraction from tunable dynamic grating. The interference of two pulses in the other words imposes a suitable grating shape for desired harmonic mode onto the plane of emitting dipoles. According to the holography, the diffracted orders preserve the wavefront of the of the incident object beam which here is the second pulse with desired transversal mode. The grating pattern by introduce a phase modulation in the dipole emission imprint such pattern in the gas medium. The Harmonics are diffracted from the IR grating and in far-field interfere. The diffracted harmonics are in phase because every point in transient grating is in phase with other point. The residual phase mismatch can be compensated using intersect angle or by wavelength detuning to change the harmonic phases.

The phase matching condition in high harmonic generation is related to the momentum conservation in the total interaction process. Here contribution of the transient grating in HHG is shown versus a two dimensional momentum geometry of the fundamental interfere beams in x-z surface. In photon picture as explained in the combination of $m_1$ photons from fundamental laser pulse with frequency $\omega_1 = \omega$ and $m_2$ photon from interfering laser pulse with frequency $\omega_2 = \omega$ for stationary grating and $\omega_2 = 2\omega$ for moving grating, contribute in high harmonic radiation with $\omega_q = q\omega$ where $q = m_1 + m_2$ and $q = m_1 + 2m_2$ for stationary and moving grating respectively. Then as shown in fig. for small intersect angles

$$\beta_{q,m_2}(\theta) = 2m_2\theta / q. \qquad (7)$$

This is exactly the same as phase modulation term in wave picture of the generation of high harmonic in the interference of two laser beam with small intersect angle [13]. The angle and consequently the phase pf the qth harmonic depends on the number of interfering photons that contribute in the generation or in other word the intensity ratio of the interfering to the fundamental beam. Consequently, the relative intensity of the interfering and fundamental beams affects the dipole amplitude. This equations represent that the divergence angle between the same order harmonics can be small when the intersect angle θ is small.

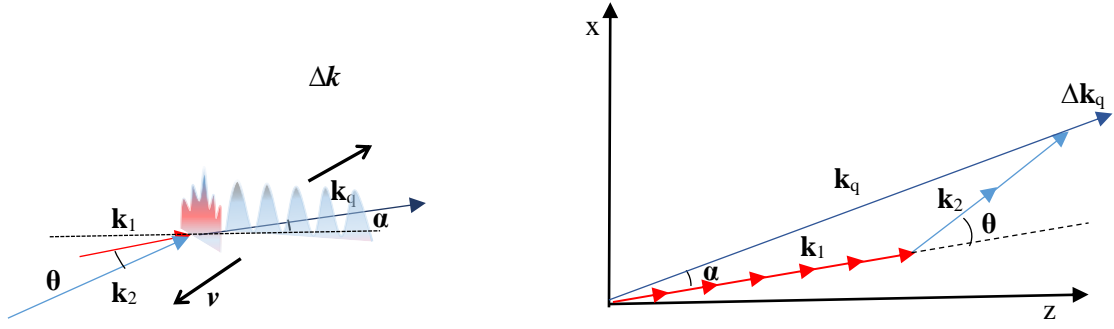

Figure 5: schematic of the fundamental and interfering momentum contribution in HHG for spatial case of sum frequency generation (SFG)

The phase mismatch in two dimensional wavevector geometry as a combination of mismatch when θ=0 and an extra geometrical offset which appeared in transversal variation of momentum on transient grating surface,

$$\Delta \mathbf{k}_q = \Delta \mathbf{k}_{\theta=0} + \Delta \mathbf{k}_{offset} \quad [13]. \qquad (8)$$

where $\Delta \mathbf{k}_{\theta=0} = \nabla \varphi_q - \frac{n_q \omega_q}{c}$. $\varphi_q$ can be obtained by Fourier transform of the induced dipole moment of interaction of atom with femtosecond pulse in strong field approximation as

$$\varphi_q = q\varphi - \varphi_d. \qquad (9)$$

The wavevector mismatch $\Delta \mathbf{k}_{\theta=0}$ is sum of the four contributions

$$\Delta \mathbf{k}_{\theta=0} = \Delta \mathbf{k}_d + \Delta \mathbf{k}_g + \Delta \mathbf{k}_n + \Delta \mathbf{k}_e. \qquad (10)$$

$\Delta k_d = \nabla \varphi_d$ denotes the wavevector mismatch induced by dipole phase which is phase change of electron wavepacket in laser field in continuum from ionization to recombination moment

$$\frac{1}{\hbar} \int_{t_i}^{t_r} dt \frac{1}{2} \left( \mathbf{P} - \mathbf{A}(t) \right)^2 \quad [14], \qquad (11)$$

where A(t) is the electric potential of the femtosecond laser pulse, can be approximated as

$$\varphi_d(q, I) = \alpha \frac{U_p}{\hbar \omega}, \qquad (12)$$

where α depends on the electron excursion time in continuum and can be calculated can be calculated based on the harmonic order and the action of the semi-classical electron trajectory from saddle point approximation [15], $U_p = I/4\omega^2$ is the ponderemotive potential with $I(\mathbf{r},t)$ as the intensity distribution of the grating.

The induced mismatch due to the modulated intensity of grating and the consequent intensity dependent phase in x-y plane can be written as

$$\Delta \mathbf{k}_d(x,z) = \frac{\alpha}{2c\varepsilon_0\omega^3}\left(\frac{\partial I}{\partial x}\mathbf{e}_x + \frac{\partial I}{\partial z}\mathbf{e}_z\right), \qquad (13)$$

where the intrinsic phase of harmonic emission can vary longitudinally and radially throughout the focus because of varied atomic response to the local laser intensity of the grating. The intensity modulation in grating pattern make an spatial variation in atomic phase. By adjusting the angle and intensity ratio the phase of harmonic as well as angle of diffraction will be changed to make constructive interference between the same order generated harmonics.

$\Delta k_g$ caused by Guoy phase (Fig. 6 shows the Guoy phase is introduced by interfering of fundamental 800 nm and interfering 400 nm pulses with intensity ratio 0.1 and intersect angle $0.1\pi$ rad.).

$\Delta k_n$ is the consequence of the dispersion in neutral gas. The refractive index of nobel gases for driving laser and generated harmonics is different ($\delta n = n - n_q$) from the sellmeier equation. Moreover the refractive index of the gas medium is proportional to the pressure (P) and because of ionization, the density of neutral atoms is also scaled by the fraction of ionization, $\eta$, giving a phase

mismatch between the driving laser and harmonic fields

$$\Delta \mathbf{k}_n(x,z) = -2\pi q P \delta n (1-\eta)/\lambda \qquad (14)$$

Finally,

$$\Delta \mathbf{k}_e(x,z) = q\, \text{Pr}_e\, N\eta\lambda \qquad (15)$$

is the consequence of the dispersion in plasma, N is the number density of atoms at 1 atm and $r_e$ is the electron radius. To generate very high harmonic orders high laser intensity is required which leads to ionization levels that are greater than the critical ionization and the harmonic intensity then becomes very low due to the large phase mismatch.

The extra offset term which caused by crossing interference for harmonic q when $m_2$ photon of interfering pulse are contributed in generation is

$$\Delta \mathbf{k}_{offset}(m_2,q,\theta) = m_2\theta^2(2m_2/q - 1)\omega/c, \qquad (16)$$

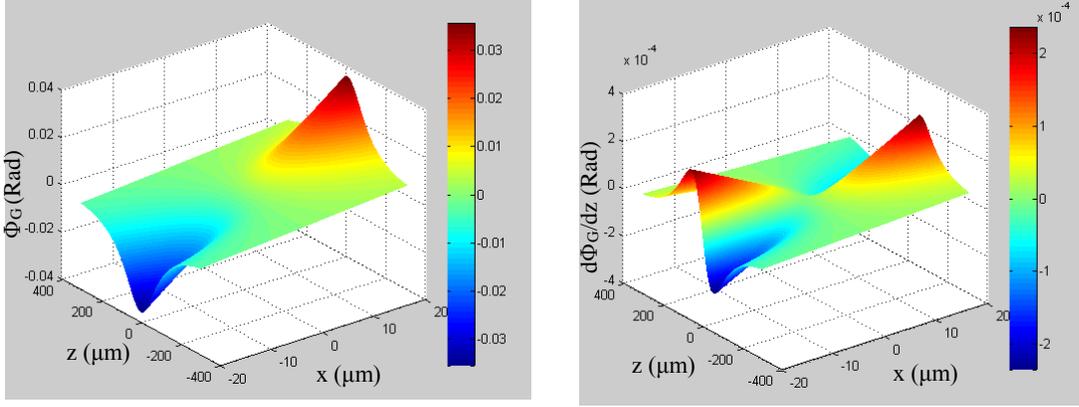

Figure 6: Guoy phase ($\Phi_G$) and its derivative respect to z for static transient grating created by two 800 nm beams with intersect angle 5°.

where c is the speed of light.

If the interfering pulse photons carry extra information such as orbital angular momentum, then from conservation of orbital angular momentum

$m_1 \mathbf{L}_1 + m_2 \mathbf{L}_2 = \mathbf{L_q}$, (17)

an additional term as a function of the intersect angle and the angular momentum of the interfering beam will be added to the wavevector mismatch

$$\Delta \mathbf{L}_{offset}(l,\theta,q) = l\theta^2 (2l/q - 1)\omega/c .$$ (18)

Here we consider the fundamental Gaussian beam with $\mathbf{L}_1 = 0$ and the Laguerre Gaussian interfering beam with angular momentum $|\mathbf{L}_2| = l$, where $|\mathbf{L_q}| = l$. It means that the vortex shape of the interfering beam transferred to the generated XUV laser beam by meaning of conservation of orbital angular momentum.

The offset terms in phase mismatch is negative and therefore can compensate for plasma dispersion mismatch by tuning the pressure, the transversal modes and the intersect angle according to the above equations.

The intersect angle as well as the intensity ratio of fundamental to interfering beam can be tuned such that having phase matched desired transversal mode respect to the transient grating pattern according to Eq. 13. The diffracted harmonic orders preserve the wavefront of the interfering beams with desired transversal mode according to the conservation of the angular momentum (Eq. 17). The generation of the helical XUV beams can be investigated using semiclassical model of HHG. The Phase of LG beam will be appeared in the phase change of the electron wave packet in Eq. (11).

The phase mismatch due to plasma dispersion and neutral dispersion according to Eq.s 14 and 15 are positive and therefore can be canceled for the negative noncollinear phase mismatch introduced in Eq.s 16 and 17 by tuning the gas pressure to scale the diffracted transverse XUV.

### 4. Simulation of the HHG based on dynamic transient grating

The high harmonic generation process using transient grating can be simulated considering multiple harmonic sources as wavelets in transversal plane that is moving with the speed of V according to Eq. 4. Each wavelet is driven by the time dependent local electric field in the pattern ceated by the interfering of the femtosecond beams with engineered transversal modes. At each point on a two dimensional dynamic grid we can calculate the dipole moment in the strong field approximation. Then, it is decomposed in the Fourier to find near field emission of harmonic orders. The far field image can be obtained by Fresnel propagation integral.

**Conclusion**

Summing over all the diffracted orders of harmonics, in fact, an overall conversion efficiency will be increased. The increase in conversion efficiency would be achieved by self-phase matching due to the modulation of geometrical parameters and also due to the phase change in generated harmonic by modulating the intensity. The method we introduced can be extended for any combination of driving lasers to produce different grating patterns with various tunable spatial properties that can be applied to create phase matched XUV pulses with engineered shapes. A control over the spin as well as orbital angular momentum of the femtosecond interfering beams is can pave the way toward tailoring more complex XUV light sources for fundamental studies and applications. In the other hand, the high harmonic efficiency of the proposed model can be tuned to be optimized by taking the advantage of the adjustable transient grating parameters which carries the spatial and temporal properties of the interfering beams. In addition by interfering of a mid-IR laser beam and its suitable harmonic combined with favorable phase matching condition it is possible to extend the application of this technique from XUV to soft x-ray with reasonable intensity for application in microscopy and study of biological systems.